\documentclass[12pt]{iopart}
\usepackage{graphicx}
\usepackage{epsfig}
\usepackage{dcolumn}
\usepackage{bm}

\def\lessim{\lower.5ex\hbox{$\; \buildrel < \over \sim \;$}}
\begin{document} \hbadness=10000
\title{Charmed Hadrons from Strangeness-rich QGP}
 \author{Inga~Kuznetsova  and Johann Rafelski}
\address{Department of Physics, University of Arizona, Tucson,
Arizona, 85721, USA}

\begin{abstract}
The yields of charmed hadrons emitted by strangeness rich QGP are
evaluated within chemical non-equilibrium statistical hadronization
model,  conserving
strangeness, charm, and entropy yields at hadronization.
\end{abstract}

\section{Introduction}
A relatively large number of hadrons containing charm and bottom
quarks are expected to be produced in heavy ion (AA) collisions at
the Large Hadrons Collider (LHC). Because of their large mass
$c,\bar c$ quarks are produced predominantly in primary
parton-parton collisions~\cite{Geiger:1993py}, at
RHIC~\cite{Cacciari:2005rk}, and thus even more so at LHC. These
heavy flavor quarks participate in the evolution of the dense QCD
matter from the beginning. In view of the recent RHIC results it can
be hoped that their momentum distribution  could reach approximate
thermalization within the dense QGP phase~\cite{vanHees:2004gq}.

We study in this report heavy quark hadronization, and
we are particularly interested in how strangeness influences
heavy and multi-heavy hadron yields (containing more than
one heavy quark). In the presence of deconfined QGP
multi-heavy hadrons are formed
from heavy quarks created in the initial NN collisions.
Therefore yields of these hadrons are expected to be  enhanced as
compared  to yields seen in single NN
collisions~\cite{Schroedter:2000ek,Becattini:2005hb}. This type of
enhancement can be considered to be an indicator of the presence of
deconfined QGP phase, for reasons which are analogous to those given
in the context of multi-strange (anti) baryon enhancement~\cite{Koch:1986ud}.
$J/\Psi (c\bar c)$  yields were for this reason
obtained in the kinetic formation and
dissociation model~\cite{Schroedter:2000ek,Thews:2005fs} but
without consideration of a strangeness content in the QGP phase.

Differing from  other recent studies   which assume that
the hadron yields after hadronization are in chemical
equilibrium~\cite{Becattini:2005hb,Andronic:2003zv},
we obtain the charm hadron yields in the statistical
hadronization approach   based on an
abundance of $u,d,s$ quark pairs fixed by the
bulk properties of a chemically equilibrated QGP phase.
Since we do not yet know to sufficient precision the
heavy flavor production at LHC or even RHIC energies,  we
present particle yields normalized by   the total charm flavor yield,
such that the particle yield results we consider are little dependent on
the  unknown total yield of charm.
The results we present were obtained for ${dN_c/ dy}\equiv c=10$.

\section{Fast Hadronization}\label{SHMsec}
We study within the statistical hadronization model (SHM)
the charmed hadron production at fixed strangeness pair yield per entropy $s/S$. 
This in general requires  the chemical non-equilibrium
version of the SHM.
The important parameters of the SHM, which control the relative
yields of particles, are the particle specific fugacity factor
${\lambda}$ and space occupancy factor ${\gamma}$. The fugacity is
related to chemical potential ${\mu} = T{\ln{\lambda}}$.
We take  ${\lambda}_i=1$ for all flavors since the small particle-antiparticle
yield asymmetry at RHIC and LHC is not relevant here.

The ratio of the number of given type particles produced, to the
number expected in chemical equilibrium is (nearly) the phase
space occupancy ${\gamma}$,  which
is the same for particles and antiparticles of the type considered.
We use occupancy factors $\gamma^\mathrm{Q}_i$ and
$\gamma^\mathrm{H}_i$ for QGP and hadronic gas phase respectively, tracking
every quark flavor ($i$ = q, s, c). For each hadron, the phase space occupancy
$\gamma^\mathrm{H}$   is the product of
$\gamma^\mathrm{H}$'s for each constituent quark. For example for
the charmed meson D($c\bar{q}$),
$\gamma^\mathrm{H}_D = \gamma^\mathrm{H}_c\gamma^\mathrm{H}_q$.
The value of $\gamma^\mathrm{H}_q$ has upper limit arising from the
condition of Bose-Einstein condensation of pions $\gamma_\pi=\gamma_q^2\le e^{m_\pi/T}$.

The number of particles of type `$i$' with mass $m_i$ per unit of rapidity is:
\begin{equation}
\frac{dN_i}{dy}={\gamma_i}n_i^{\rm eq}\frac{dV}{dy}.  \label{dist}
\end{equation}
Here $dV/dy$ is system volume associated with the unit of rapidity,
and $n_i^{\rm eq}$ is a Boltzmann particle density in chemical equilibrium:
\begin{equation}\label{BolzDis}
n_i^{\rm eq} = g_i\int\frac{d^3p}{(2\pi)^3}\lambda_i\exp(-\sqrt{p^2+m_i^2}/T)
= \lambda_i\frac{T^3}{2\pi^2}g_i(m_i/T)^2K_2(m_i/T). \label{distapr}
\end{equation}

During a fast transition between QGP and HG phases  strange and heavier
quark flavor yields are preserved, as are the entropy per unit
of rapidity~\cite{Bjorken:1982qr},  and the specific, per rapidity,
hadronization volume:
\begin{equation} \label{Sflcons}
\frac{dN^\mathrm{H}_i}{dy}=\frac{dN^\mathrm{Q}_i}{dy}=\frac{dN_i}{dy},
\ i=s,c;\quad
\frac{dS^\mathrm{H} }{dy}=\frac{dS^\mathrm{Q} }{dy}=\frac{dS}{dy};
\quad
\frac{dV^\mathrm{Q}}{dy}=\frac{dV^\mathrm{H}}{dy}.
\end{equation}
The yields of  hadrons after hadronization are given by
Eq.\,(\ref{dist}), the three unknown
$\gamma^\mathrm{H}_q,\,\gamma^\mathrm{H}_s$ and
$\gamma^\mathrm{H}_c$ can be determined from their values in the QGP
phase, or equivalently corresponding flavor yields, i.e.
$\gamma^\mathrm{Q}_i$ or $dN^Q_i/dy$ given in Eq.\,(\ref{Sflcons}).
$\gamma_i\ne 1$ implies that hadron yields are in general not in a
chemical equilibrium. We will show how this influences the relative yields of heavy flavored particles in the final state.

The number of strange quark pairs determines the value of $\gamma^{\mathrm{H}}_{s}$:
\begin{equation}
\frac{dN_{s}}{dy}=\frac{dV}{dy}\left[{{\gamma^{\mathrm{H}}_{s}}}
  \left(\gamma^{\mathrm{H}}_qn^{\rm eq}_{K}+
      \gamma^{\mathrm{H}\,2}_qn^{\mathrm{eq}}_Y \right)+
      \gamma^{\mathrm{H}\,2}_s(2\gamma^{\mathrm{H}}_qn^{\rm eq}_\Xi+ n^{\mathrm{eq}}_{s\,{\rm hid}})
+ 3\gamma^{\mathrm{H}\,3}_sn^{\rm eq}_\Omega\right], \label{gammas}
\end{equation}
where $n^{\mathrm{eq}}_i$ are densities of strange
mesons and baryons calculated  using Eq.\,(\ref{dist}) in chemical equilibrium,
$n^{\mathrm{eq}}_{s\,{\rm hid}}=n^{\rm eq}_\phi+P_sn^{\rm eq}_\eta$ and $P_s$
is the strangeness content of the $\eta$.
The pattern of this  calculation follows an
established approach by using SHARE 1.2 program~\cite{Torrieri:2004zz} in the calculation.

For charm hadrons $\gamma^{\mathrm{H}}_c$ is obtained from:
\begin{equation}
\frac{dN_{c}}{dy}=\frac{dV}{dy}\left[{\gamma^{\mathrm{H}}_{c}}n^{c}_{\mathrm{op}}
+
\gamma^{\mathrm{H}\,2}_{c}(n^{\rm eq}_{c\,\mathrm{hid}}
+
2\gamma^{\mathrm{H}}_qn^{\mathrm{eq}}_{ccq}
+
2\gamma^{\mathrm{H}}_sn^{\mathrm{eq}}_{ccs})\right]; \label{gammacb}
\end{equation}
where open `op'  charm yield is:
\begin{equation}
n^{c}_{\mathrm{op}}\!=\gamma^{\mathrm{H}}_qn^{\mathrm eq}_{D}+\gamma^{\mathrm H}_sn^{\mathrm eq}_{Ds}+
{\gamma^{\mathrm{H}\,2}_q}\,n^{\rm eq}_{qqc}+{\gamma^{\mathrm{H}}_s}{\gamma^{H}_q}n^{\mathrm eq}_{sqc}+
{\gamma^{\mathrm{H}\,2}_s}\,n^{\mathrm{eq}}_{ssc};
\end{equation}
Here $n^{\mathrm{eq}}_{D}$ and $n^{\mathrm{eq}}_{Ds}$ are densities of
$D$ and $D_s$ mesons respectively in chemical equilibrium, $n^{\mathrm{eq}}_{qqc}$ is
equilibrium density of baryons with one charm and two light quarks,
$n^{\mathrm{eq}}_{qsc}$ is density of baryons with one light, one $s$ and one $c$ quarks,
$n^{\mathrm{eq}}_{ssc}$ is density of baryons  with one
charm and two strange quarks ($\Omega^0_c(ssc)$) and
$n^{\rm eq}_{c\,\mathrm{hid}}$ is the density of particles with both a
charm quark and an anticharm quark (C=0, S=0).

The use of the hadron phase space (denoted by H above) does not
imply the presence of a real physical `hadron matter' phase: the
SHM particle yields will be attained  solely on the basis of
availability of this  phase space. In fast hadronization
one can assume that there are  practically only free-streaming particles
in the final state.

Thinking in these terms, one can imagine that especially
for heavy quark hadrons some particles are
pre-formed in the deconfined plasma, and thus the heavy hadron yields
may be based on a value of temperature which is higher than the global
value expected  for other hadrons.
For this reason we will study in this work a rather wide range $140<T<260$ MeV.

We will use as a reference a QGP state with the ratio strangeness (s)
to entropy (S) fixed. We take it  be in the range  $s/S=$0.03 -- 0.04, and consider 
$dV/dy=$600 -- 800${\mathrm{fm}}^{3}$ at $T=200\,{\mathrm{MeV}}$. This
corresponds to total particles' multiplicity of about 4000--5000 after
hadronization for T=140 MeV, according to calculations using SHARE
1.2~\cite{Torrieri:2004zz}.  This is within the expected  range of total
hadron multiplicity per unit of rapidity  for LHC. 
We compare our results to the `benchmark' reference yield which
is obtained using
$\gamma^{\mathrm{H}}_s=\gamma^{\mathrm{H}}_q=1$
(chemical equilibrium). 

\section{Yields of charmed hadrons}\label{heavyFlSec}

\subsection{D, $D_s$ meson yields}\label{cbMesYielSec}
\begin{figure}
\centering
\includegraphics[width=8.6cm,height=11.5cm]{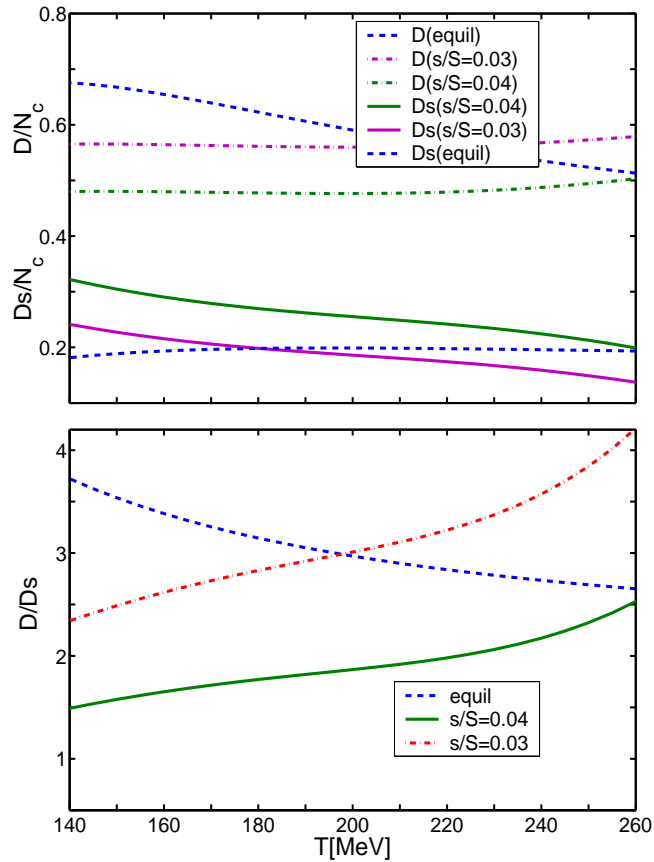}
\caption{\small{The upper panel:  charm meson yield per total charm yield $N_c$.
Equilibrium (dashed lines, blue ), non-equilibrium (solid and dash-dot
lines are for $D_s$ and $D$, respectively). 
The  lines for $s/S=0.04$ (green) are more centrally located in each frame,
outer (purple) lines are for $s/S=0.03$. The lower panel: ratios
$D/D_s$ as a function of T. Solid line is for $s/S=0.04$, dash-dot
line is for $s/S=0.03$, dashed line is for
$\gamma^{\mathrm{H}}_s=\gamma_s^{\mathrm{H}}=1$
 (chemical equilibrium).} } \label{Dmes}
\end{figure}

Considering Eq.\,(\ref{dist}), and using $\gamma^{H}_c$,
$\gamma^{H}_s$, $\gamma^{H}_q$ at a given $T$  we have now all the
inputs required to compute relative particle yields of all charmed
hadrons. In figure
\ref{Dmes} we show in the upper panel the fractional yields of
non-strange $D/N_c$ and strange $D_s/N_c$ charmed
mesons normalized by the total number
of charm quarks $N_c$ as functions of hadronization temperature. The
dashed lines are for chemical equilibrium. The extreme upper and
lower lines are for $s/S=0.03$ (purple, solid and dash-dot lines),
while the central  lines are for $s/S=0.04$ (green, solid and
dash-dot lines). We recall that the chemical equilibrium of
strangeness in QGP corresponds to  $s/S\simeq 0.03$. The yield of $D$
is the sum over all $(\bar q  c)$ states i.e. $D^0$ and $D^+$ ($D^+
\approx D^0$) mesons and resonances. Similarly,  $D_s(\bar s c)$ is
the sum over all resonances  of $D_s$ mesons.

In the lower panel we show the ratios $D/D_s$ as a function of
hadronization temperature. 
There is a considerable deviation in the ratios obtained for a fixed $s/S$ 
from the chemical equilibrium results, except for the 
accidental values of $T$ where
the equilibrium results (dashed line) cross 
the fixed $s/S$ results. While the chemical equilibrium
results show significant difference between strange and non-strange
heavy mesons, for the fixed $s/S$  case, with increasing $s/S$ and towards
low $T$ these yields become nearly equal. If the heavy meson
yields are established at temperatures similar to regular hadrons,
relative enhancement by factor 2--2.5 in strange-heavy mesons is to
be expected for LHC. For RHIC conditions ($s/S=0.03$, $c=\bar{c}=2$)
$D/D_s$ ratio is lower than the chemical equilibrium value for $T<0.2$.
This ratio decreases by factor of 1.6 for $T\to 140$ MeV.

\subsection{Yields of hadrons with two heavy quarks}\label{MultiSec}
\begin{figure}[!t]
\centering
\includegraphics[width=8.6cm,height=11.5cm]{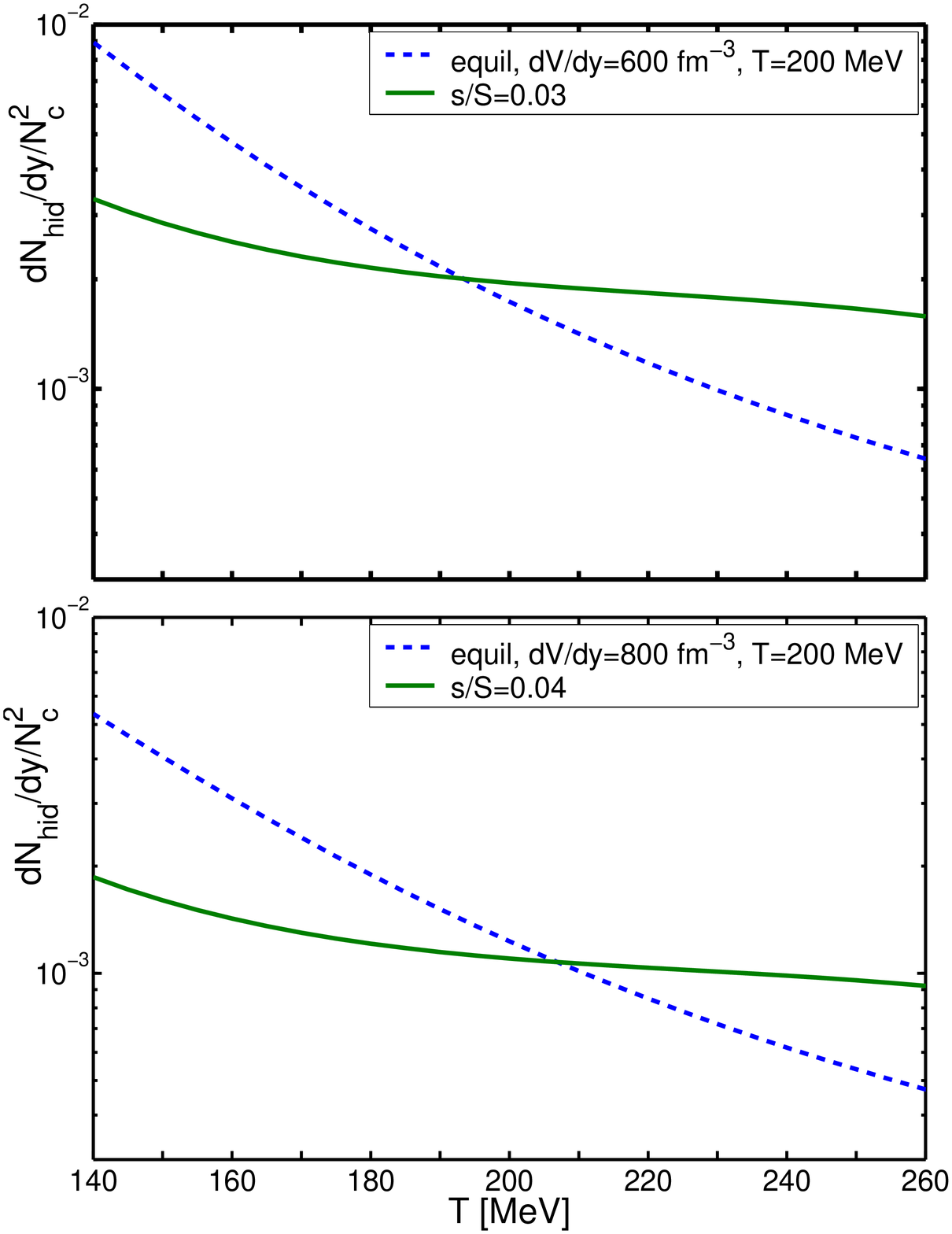}
\caption{\small{$c\bar{c}/N^2_c$ relative yields as a function of
hadronization temperature $T$ for $s/S=0.03$ with reference volume
$dV/dy=600$\,fm$^{-3}$ at  $T=200$ MeV (upper panel); and
$s/S=0.04$ with $dV/dy=800\,\mathrm{fm^{-3}}$ at  $T=200$ MeV(lower panel). 
Dashed lines are for chemical equilibrium.}} \label{cc}
\end{figure}

The hadron yields with two heavy quarks  are  model dependent, 
being proportional to $1/dV/dy$ because $\gamma^H_{i}$ for heavy
quarks is proportional to $1/dV/dy$. Thus our result is dependent on
the reaction volume, or on the total assumed charm yields.
In figure~\ref{cc} we show the yield of hidden charm
$c\bar{c}$ (sum over all states of $c\bar{c}$) mesons normalized by the square
of charm multiplicity $N_c^2$ as a
function of hadronization temperature $T$
for two different reference values of 
volume at  $T = 200$ MeV: 
$dV/dy=600$\,fm$^3$ (upper panel), and
$dV/dy=800$\,fm$^3$ (lower panel). We consider again cases with
$s/S=0.03$ (upper panel, solid line) and $s/S=0.04$ (lower panel,
solid line). For comparison, 
the chemical equilibrium $c\bar{c}$ mesons yields are
shown (dashed lines on both panels). 
Entropy per unit of rapidity $dS/dy$ is conserved, during 
cooling/expansion of the QGP,  which 
implies  $VT^3\simeq$Const.  In the second case considered
rapidity density of strangeness is 78\% higher than in the first. 

The yield of $c\bar{c}$ mesons is much smaller at fixed $s/S=0.04$ than
in chemical equilibrium for the same $dV/dy$ for large range of hadronization
temperatures. For $s/S=0.03$
the effect is similar, but charmonium suppression is slightly less pronounced.
This new mechanism of charmonium suppression occurs
due to competition with the yield of strange-heavy mesons. The
enhanced yield of $D_s$ in effect depletes the pool of available
charmed quark pairs, and fewer hidden charm $c\bar{c}$ mesons are formed.
For particles with two heavy quarks the
effect is larger than for hadrons with one heavy quark and light
quark(s).
\section{Conclusions}\label{concSec}

We have considered here  the abundances
of heavy flavor hadrons within the statistical hadronization model.
While we compare the yields to the expectations based on
chemical equilibrium yields of light and strange quark pairs,
we present results based on the hypothesis that
the QGP entropy and QGP flavor yields determine
the values of phase space occupancy $\gamma^\mathrm{H}_i$, $i=q,s,c $,
which are of direct interest in study of the heavy hadron yields.

We studied how the (relative) yields of strange and non-strange
charmed mesons vary with strangeness content. For a chemically
equilibrated QGP source, there is considerable shift of the yield
from non-strange $D$ to the strange $D_s$. Since the expected
fractional yield $D_s/N_c\simeq 0.2$ when one assumes
$\gamma^{\mathrm{H}}_s=\gamma^{\mathrm{H}}_q=1$, one should be able
to falsify chemical (non-)equilibrium hypothesis easily: the
expected enhancement of the strange heavy mesons being at the level
of 40\% when $s/S=0.03$, and yet greater, when greater strangeness
yield is available.

Another  consequence of this result is that we find a relative suppression of the
multi-heavy hadrons, except when they contain strangeness. The somewhat
ironic situation is that while higher charm QGP yield enhances production
of $c\bar c$ states, the enhanced strangeness suppresses this effect.

\vspace*{.2cm}
\subsubsection*{Acknowledgments}
Work supported by a grant from: the U.S. Department of Energy  DE-FG02-04ER4131.
IK thanks SQM2006 for local support.

%
\vspace*{-0.3cm}
\section{References}

\end{document}